\documentclass{article}

\usepackage{PRIMEarxiv}

\usepackage[utf8]{inputenc} 
\usepackage[T1]{fontenc}    
\usepackage{hyperref}       
\usepackage{url}            
\usepackage{booktabs}       
\usepackage{amsfonts}       
\usepackage{nicefrac}       
\usepackage{microtype}      
\usepackage{lipsum}

\usepackage{tabularx}
\usepackage{fancyhdr}       
\usepackage{graphicx}       
\graphicspath{{media/}}     
\usepackage[square,numbers]{natbib}
\title{Impacts of Data Preprocessing and Hyperparameter Optimization on the Performance of Machine Learning Models Applied to Intrusion Detection Systems}

\author{
  Mateus Guimarães Lima, Antony Carvalho, João Gabriel Álvares \\
  Command and Control Company \\
  Brazillian Army \\
  Brasília - DF - Brasil\\
  \texttt{\{lima.mateus,antony.carvalho,alvares.joao\}@eb.mil.br} \\
   \And
  Clayton Escouper das Chagas, Ronaldo Ribeiro Goldschmidt \\
  Instituto Militar de Engenharia (IME/RJ) \\
  Rio de Janeiro - RJ - Brasil \\
  \texttt{escouper@ime.eb.br} \\
}

\begin{document} 

\maketitle

\begin{abstract}

In the context of cybersecurity of modern communications networks, Intrusion Detection Systems (IDS) have been continuously improved, many of them incorporating machine learning (ML) techniques to identify threats. Although there are researches focused on the study of these techniques applied to IDS, the state-of-the-art lacks works concentrated exclusively on the evaluation of the impacts of data pre-processing actions and the optimization of the values of the hyperparameters of the ML algorithms in the construction of the models of threat identification. This article aims to present a study that fills this research gap. For that, experiments were carried out with two data sets, comparing attack scenarios with variations of pre-processing techniques and optimization of hyperparameters. The results confirm that the proper application of these techniques, in general, makes the generated classification models more robust and greatly reduces the execution times of these models' training and testing processes.

\end{abstract}

\section{Introduction}

Communication networks play a vital role in modern society, being present in numerous locations, from corporate environments to households. As a consequence of this scenario, protection against cyber threats is a crucial concern, as communication networks are susceptible to sophisticated attacks that seek to compromise the integrity, confidentiality, and availability of transmitted information \cite{angin2020blockchain}. Therefore, effective intrusion detection in these networks is fundamental for safeguarding transmitted data.

Intrusion Detection Systems (IDS), are traffic monitoring solutions that identify suspicious activities in the network or hosts, such as copying, modifying, or deleting applications, files, or directories. IDSs are based on signatures, anomalies, or, more recently, on anomaly detection with machine learning (ML) \cite{shaukat2020performance}.

The ability to detect advanced threats in real-time using traditional intrusion detection techniques is limited, which is why ML has become an important tool for enhancing these techniques. ML offers a sophisticated and adaptive approach to intrusion detection, given its effectiveness against Zero-Day threats \cite{zhou2019evaluation}. The use of ML algorithms can enable IDSs to be trained to recognize suspicious network traffic patterns, identify anomalous behaviors, and thus, detect threats that are difficult to detect by other methods.

Many studies have explored the incorporation of ML techniques into IDSs, in which, during their flow, pre-processing actions and optimization of hyperparameters are executed to prepare the data for ML algorithm processing. Although there is consensus that data pre-processing and hyperparameter optimization are important steps in the ML flow, existing works in the literature generally limit themselves to reporting the pre-processing actions performed and the use of optimized hyperparameters, without further investigating the impacts of each action on the generated models.

Within the presented context, this work aims to answer the following research question: what impacts can pre-processing actions and hyperparameter optimization of ML algorithms have on the performance of classification models for threat identification in communication networks and also on the execution time of training and testing processes of these models?

For this purpose, binary classification experiments (between normal traffic and cyber-attacks) were conducted with two widely circulated datasets in the cybersecurity research community, and different experimentation scenarios were compared. As a result, it was found that the use of the proposed pre-processing actions, combined with the optimization of hyperparameter values, achieved the best result in the calculated performance metrics, in most of the evaluated attack scenarios and algorithms. Furthermore, pre-processing significantly improved the execution time of both training and testing, while maintaining or improving predictive performance metrics. It is noteworthy that the reduction in training time is a relevant factor in the current reality, as the increasing amount of new data may require continuous updating of ML-generated models. Moreover, reducing testing times can lead to models that, when deployed on the equipment they will be embedded in, make intrusion identification faster, which is critical especially in real-world scenarios and with sensitive data.

\section{Related Works}\label{sec:trabrel}
Google Scholar searches were conducted for articles using the terms "machine learning", "preprocessing", and "intrusion detection system". From the search results, publications most related to the present study were selected.

Data preprocessing typically involves three main stages: cleaning, normalization, and feature selection \cite{kasongo2020performance}. According to \cite{ullah2022hdl}, the primary goal of feature selection is to avoid overfitting and underfitting, as well as to improve performance and reduce model training and response time. In cleaning, non-numeric or symbolic data is generally removed or replaced, as it plays a less relevant role in intrusion detection in certain datasets \cite{halimaa2019machine}.

The study described in \cite{umar2020effects} indicates that the advent of larger datasets has had negative impacts on the performance and computational time of ML-based IDS. To overcome such issues, many researchers have utilized techniques such as feature selection and normalization in data preprocessing. However, the authors conclude that normalization may be more important than feature selection for improving performance and computational time associated with model generation and application.

The work \cite{liu2019machine} highlights that most studies emphasize detection results and often employ complex models and extensive data preprocessing methods, leading to low efficiency. To mitigate damage as much as possible, however, IDSs need to detect attacks in real-time. Thus, there is a trade-off between effectiveness and efficiency.

In \cite{farhan2020performance}, it is recommended that works utilize some feature selection methods to increase accuracy and detection rate, as well as reduce false alarms and minimize computation time. Additionally, it is pointed out that optimizing hyperparameter values is recommended for greater efficiency.

\begin{table}[ht]
\caption{Summary of related works}
\label{tab:trab_rel}
\centering
\small
\begin{tabularx}{\textwidth}{lXXXXXX}
\hline
\textbf{Reference} & \textbf{Algorithm} & \textbf{Technique} & \textbf{Dataset} & \textbf{Evaluated preprocessing impact?} & \textbf{Evaluated hyperparameter optimization impact?} & \textbf{Evaluation of execution time (training and testing)} \\
\hline
\cite{ahmad2019data} & NB, SVM, and KNN & Normalization, discretization, CFS, PSO & Kyoto 2006, KDD Cup99, UNSW-NB15 & Yes & No & No \\
\cite{kumar2016network} & NB, J48, REPTree & CFS, IGF, GRF & NSL-KDD & Yes & No & Yes \\
\cite{singh2019evaluation} & NB, DT, J48 & Subset Attribute Evaluator, Classifier Subset Evaluator & CICIDS-2017 & Yes & No & Yes \\
\cite{farhan2020performance} & DNN & Manual feature selection, normalization & CSE-CIC-IDS 2018 & Yes & No & No \\
\cite{ullah2022hdl} & DNN & Data cleaning, randomization, normalization, ETC, SMOTE & CIC DoS 2016, CICIDS 2017, CSE-CIC-IDS 2018 & Yes & No & Yes \\
\cite{zhou2019evaluation} & RF, NB, DT, MLP, KNN, quadratic discriminant analysis & Data cleaning, formatting, real number reduction & CIC-AWS-2018 & Yes & No & Yes \\
\cite{aslan2022using} & Bagging, J48, RandomTree, AdaBoost, BN & CFS & NSL-KDD, CIC-DDoS2019 & Yes & No & No \\
\cite{halimaa2019machine} & SVM, NB & Nominal attribute conversion, non-numeric attribute removal, randomization, attribute reduction with CfsSubsetEval() & KDD Cup 1999 & Yes & No & No \\
\cite{kasongo2020performance} & KNN, LR, ANN, SVM, DT & Normalization, XGBoost-inspired filter-based method for attribute importance scores & UNSW-NB15 Dataset & Yes & No & No \\
\hline
\end{tabularx}
\end{table}

Table \ref{tab:trab_rel} summarizes the related works. Upon analysis, the variety of approaches used is evident, i.e., there is no consensus on which preprocessing techniques and hyperparameter optimization lead to better results. The works superficially addressed the impacts of the techniques on ML models, focusing on the algorithms of the learning flow and their association with performance, overlooking the details of the impact of preprocessing and hyperparameter optimization on model robustness regarding intrusion detection capability and time performance, reinforcing the gap presented in the addressed research question.

\section{Materials and Methods}\label{sec:matmet}

The methodology evaluates the impact of data preprocessing and hyperparameter optimization on predictive performance and execution times of ML models for detecting intrusions in networks. Initially, the computational environment and datasets used are presented. Then, the workflow is detailed.

\subsection{Work Environment}\label{sec:amb}

The testing and training environment for the algorithms was virtualized on a server running Proxmox Virtual Environment version 6.3-3. A virtual machine was created with the following specifications: Intel(R) Xeon(R) Silver 4214R 2.40GHz CPU with 48 cores (4 sockets of 12 cores), 128 GiB RAM, and Linux Mint version 21.1 operating system. The programming language used was Python 3.10.6 with libraries including Scikit-Learn version 1.0.2, Pandas version 2.0.0, Numpy version 1.23.5, Tensorflow version 2.12.0, and XGBoost version 1.7.5.

The metric values and execution times of all algorithms on the datasets used in this study, as well as the scripts and programs for automating preprocessing and training and testing, can be accessed in the project repository: \url{https://github.com/comp-ime-eb-br/ml-4-cyber-def}

\subsection{Dataset}\label{sec:dataset}

The limited availability of data on cyber-attacks in real networks is a consequence of privacy and security protection. Many datasets used to train anomaly detection algorithms cannot be shared or do not reflect real attacks and new techniques. This study used two well-known datasets: the CSE-CIC-IDS2018 and the KDD Cup 1999.

The CSE-CIC-IDS2018 dataset was developed to overcome limitations in the analysis of intrusion detection systems by creating abstract user profiles that represent events and behaviors observed in the network. It includes seven different attack scenarios and network traffic information and logs from victim machines, which in this article were evaluated separately using the workflow methodology presented in the following section. The KDD Cup 1999 is a dataset based on the DARPA Intrusion Detection Evaluation Program of 1998, prepared to evaluate research in intrusion detection in defense networks.

Both datasets are valuable for the purpose of this study, providing realistic data on cyber-attacks. More information about the datasets used can be found at the links \url{https://registry.opendata.aws/cse-cic-ids2018} and \url{https://kdd.org/kdd-cup/view/kdd-cup-1999}.

\subsection{Workflow}\label{sec:wf}

\begin{figure}[ht]
\centering
\includegraphics[width=0.85\textwidth]{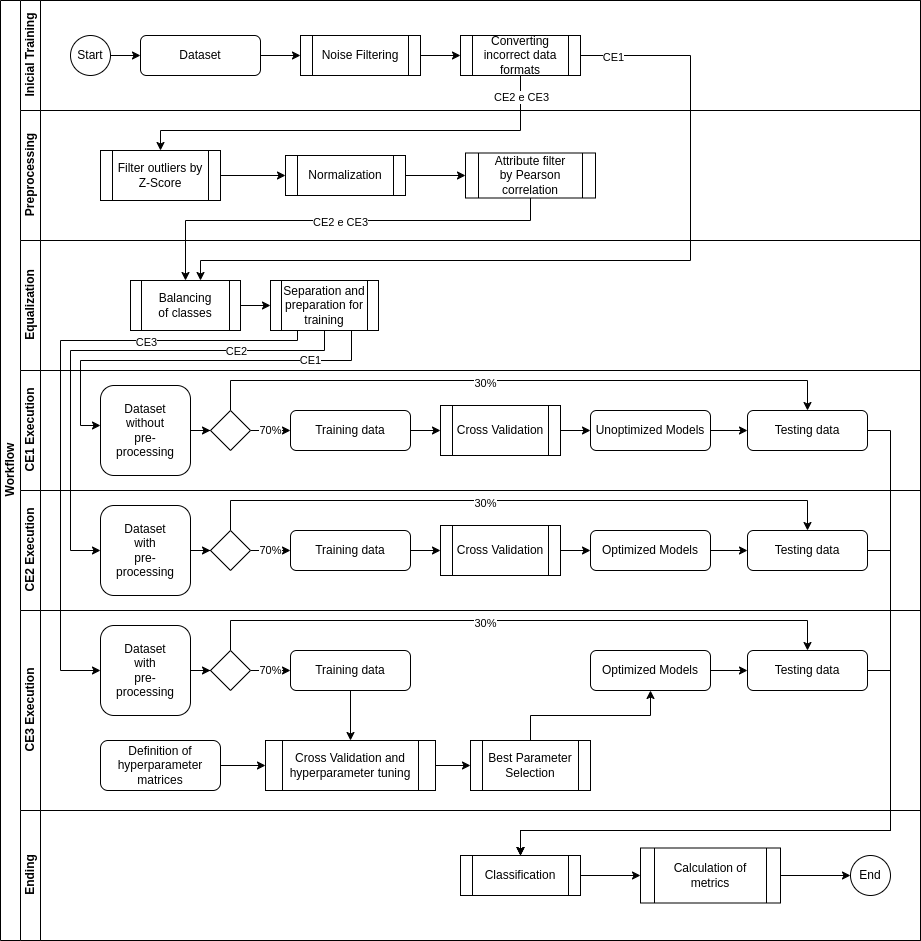}
\caption{Workflow}
\label{fig:workflow}
\end{figure}

Figure \ref{fig:workflow} illustrates the overall flow of the proposed methodology, applied to each of the datasets within the indicated separations. The following subsections provide a more detailed explanation of this flowchart.

\subsection{Experimentation Scenarios}\label{sec:ce}

Three experimentation scenarios were analyzed as comparison experiments: CE1, CE2, and CE3 (Table \ref{tab:cenarios}). Each analysis was conducted separately, with the same metrics and cross-validation methods on each dataset. The comparison allows identifying the influence of each step in obtaining high-quality models and in the performance of algorithm testing and training times, to guide answers to the proposed research question.

\begin{enumerate}
\item \textbf{CE1}: Analysis without preprocessing and without hyperparameter optimization. The datasets underwent only initial treatment, and no refinement of hyperparameter default values was applied.
\item \textbf{CE2}: Analysis with preprocessing and without hyperparameter optimization. Preprocessing was applied (as per Section \ref{sec:preproc}), but no training refinement was performed, maintaining default hyperparameters. In this scenario, the impact of preprocessing on the execution time of algorithm testing and training was evaluated.
\item \textbf{CE3}: Analysis with preprocessing and with hyperparameter optimization. Preprocessing was applied to the datasets, and optimization of hyperparameter values was performed (as per Section \ref{sec:hpevc}). This scenario aims to obtain optimized results, allowing measurement of the impact of preprocessing and hyperparameter refinement on the final prediction.
\end{enumerate}

\begin{table}[ht]
\caption{Comparison Experiments}
\label{tab:cenarios}
\centering
\begin{tabular}{|c|c|c|}
\hline
\textbf{}      & \textbf{Preprocessing} & \textbf{Hyperparameterization} \\
\hline
\textbf{CE1}         & $\times$            & Default            \\
\hline
\textbf{CE2}        & $\checkmark$            & Default            \\
\hline
\textbf{CE3}        & $\checkmark$            & Optimized            \\
\hline
\end{tabular}
\end{table}

\subsubsection{Initial Treatment} \label{sec:tratamento}
Data quality is essential for the success of data science projects. Raw data often contains noise, inconsistencies, missing values, and outliers, which can impair the effectiveness of ML algorithms. Through appropriate filtering and preprocessing, however, cleaner and higher-quality data are generated for algorithm training \cite{cousineau2010outliers}. Dataset preparation was carried out in three main stages. In the first stage, initial treatment was performed, including cleaning and conversion of incorrect data formats (subsection \ref{sec:tratamento}). In the second stage, more detailed data preprocessing was performed (subsection \ref{sec:preproc}). The last stage (subsection \ref{sec:equalizacao}) consisted of class balancing and separation by types of attacks, which was only performed on CSE-CIC-IDS2018, as the number of samples of some attacks in KDD Cup 1999 is very small, making this separation unfeasible.

In the first execution stage (first horizontal lane of Figure \ref{fig:workflow}), cleaning and elimination of attributes \cite{kasongo2020performance} that were different among datasets were performed to standardize the analysis (Figure \ref{fig:workflow}, box "Noise filtering"). Also, incorrect categorized values in the datasets were converted (some numeric values were identified as characters, objects, or strings by the program), and the Timestamp attribute was converted to a numeric value with Unix Epoch standard for better analysis by the algorithms (Figure \ref{fig:workflow}, box "Converting incorrect data formats").

\subsubsection{Preprocessing Actions}\label{sec:preproc}

In the second stage (second lane of Figure \ref{fig:workflow}), outlier values were filtered out to eliminate possible measurement errors or extrapolations in data readings or simply data that were very disparate from the others, according to the decision threshold used. The z-score value filter methodology was used \cite{cousineau2010outliers}, with $threshold=7$. Next, dataset normalization was performed \cite{kasongo2020performance} according to Equations \ref{eq1} and \ref{eq2}:

\begin{equation}
\label{eq1}
X_{std} = (X - X.min(axis=0)) / (X.max(axis=0) - X.min(axis=0))
\end{equation}
\begin{equation}
\label{eq2}
X_{scaled} = X_{std} * (max - min) + min
\end{equation}

The minimum (\textit{min}) and maximum (\textit{max}) values used were 0 and 1, respectively, for all analyzed attributes to make them more robust to very small standard deviations of attributes and to optimize computational power. By normalizing the dataset while maintaining the correlation between the data, the classification remains coherent, and the testing and training time is reduced \cite{weijun2011method}.

Still, in the second stage, a correlation filter was applied to select the most relevant attributes for the ML models \cite{ahmad2019data}. In this study, this filter was applied with the objectives of eliminating multicollinearity (which occurs when two or more attributes are highly correlated, potentially negatively affecting the model's performance) and increasing predictive performance by removing irrelevant or redundant attributes, making the model focus on more informative attributes relevant to the problem at hand.

The method for performing the filtering was the Pearson correlation filter \cite{sugianela2020pearson}, a technique used to assess the linear relationship between two variables, which can have a positive (joint increase in variables) or negative (inversely proportional variables) relationship, from 1 to -1. The decision threshold value adopted was $threshold = 0.99$, where attribute correlation is performed by pairwise comparison of attributes with the threshold, and if the pair has a correlation above the threshold, only one of the two attributes is eliminated to avoid information loss.

\subsubsection{Equalization}\label{sec:equalizacao}

In the third stage (third lane), dataset balancing was performed by equalizing the attributes followed by separation by attacks. The CSE-CIC-IDS2018 data were separated by days, with each day representing one or more different attacks. Some of these data were extremely imbalanced, which reduced the training efficiency of the models by causing bias and overfitting. Balancing was performed to address these issues. No balancing was performed for the DoS Slowloris and DDoS LOIC UDP attacks due to the low number of samples compared to benign traffic on the days of these attacks. For these attacks, the classifiers were evaluated for their ability to detect anomalies with a disproportionately larger number of benign samples.

To accurately analyze the performance of the algorithms against different types of malicious techniques to discover which could be classified with higher or lower accuracy, separation by types of attacks was performed, namely: BotNet, DDoS HOIC, DDoS LOIC HTTP, DDoS LOIC UDP, DoS GoldenEye, DoS Hulk, DoS SlowHTTPTest, DoS Slowloris, FTP BruteForce, Infilteration, SSH BruteForce, and BruteForce Web XSS.

The KDD-1999 dataset was not separated into attacks due to the low number of samples in some of these attacks, which would cause difficulty in balancing between attributes. Instead, KDD-1999 was tested as a whole.

\subsubsection{Partitioning}\label{sec:particionamento}

In the execution lanes of the experimentation scenarios in Figure \ref{fig:workflow}, the dataset was partitioned into two main sets, training and testing, in a stratified manner by class, using the \textit{sklearn.model\_selection.train\_test\_split} function. This partitioning was the same with and without preprocessing to maintain standardization in the evaluation. The training set, responsible for feeding the models during the learning process, consisted of 70\% of the data. The test set, which was not used during training, represented the remaining 30\% of the data.

\subsubsection{Cross-Validation and Hyperparameter Optimization}\label{sec:hpevc}

Also in the execution lanes of the experimentation scenarios in Figure \ref{fig:workflow}, parameter refinement was performed with a hyperparameterization technique that involved analyzing different combinations of hyperparameters for each selected algorithm.

To ensure the robustness of the results and avoid overfitting on the training dataset, the technique of $K$-fold cross-validation (in this study, $K = 5$) was applied, which involves partitioning the training data into $K$ equal-sized subsets and using each subset as the validation set while the others act as training sets, repeating this process iteratively until all subsets have been used as validation sets \cite{5342427}. The technique was applied to adjust algorithm hyperparameters more reliably and obtain a more accurate estimate of the model's performance.

The grid search technique, implemented through the \textit{sklearn.model\_selection.GridSearchCV} function, was employed along with cross-validation to automate the search for the combination of hyperparameter values that maximized the models' performance.

\subsubsection{Classification Algorithms}\label{sec:algoritmos}

Based on the nature of the anomaly detection problem and the characteristics identified in the data preprocessing, the following machine learning algorithms were selected (closing lane in Figure 1):
\begin{enumerate}
    \item \textit{Random Forest}: The implementation of the algorithm used was the one provided by the \textit{sklearn.ensemble.RandomForestClassifier()} function, based on \cite{geurts2006extremely}.
    
    \item \textit{Decision Tree}: The \textit{sklearn.tree.DecisionTreeClassifier()} function was used, which implements a simple decision tree classifier.
    
    \item \textit{XGBoost}: Implementation in the XGBoost library, incorporates machine learning algorithms into the Gradient Boosting framework, using a parallel tree boosting method (also called GBDT, GBM).
    
    \item \textit{Naive Bayes}: Implemented with the \textit{sklearn.naive\_bayes.GaussianNB()} function, an implementation of the Naive Bayes classifier in the scikit-learn library that assumes the input attributes follow a Gaussian (normal) distribution.
    
    \item Neural Network: A neural network built with the Keras library was used, with techniques to optimize the training of the neural network, such as using the Adam optimizer, an early stopping callback, and binary cross-entropy loss function. The evaluation metric adopted was binary accuracy. The architecture of the neural network used was as follows:
    \begin{enumerate}
        \item Input Layer: Receives the input data of the network with a number of neurons that varies according to the number of attributes of the input dataset.
        \item Batch Normalization Layers: Normalize the input data, helping to stabilize the training of the network and speed up convergence.
        \item Dense Layers: Contain neurons connected to all neurons of the previous layer. There were three dense layers, with 128, 64, and 32 neurons, respectively. Each dense layer used the ReLU activation function, which introduces non-linearity into the network.
        \item Dropout Layers: Help prevent overfitting by randomly deactivating some neurons during training. A dropout rate of 0.3 was set (based on empirical tests with values between 0 and 1), which means that 30\% of the neurons were deactivated at each parameter update.
        \item Output Layer: Produces the final output of the network with the Sigmoid activation function, providing a continuous value between 0 and 1, which represents the probability of belonging to one of the classes.
    \end{enumerate}
\end{enumerate}

\subsubsection{Evaluation Metrics}\label{sec:metricas}

The predictive performance of the ML algorithms was evaluated (closing lane in Figure 1) based on the following metrics:
\begin{itemize}
  \item Accuracy: Measures the proportion of instances correctly classified relative to the total number of instances.
  \item Precision: Indicates the proportion of instances correctly classified as anomalies relative to the total number of instances classified as anomalies. The weighted average of this metric was calculated for each class in the datasets to consolidate the results.
  \item Recall: Represents the proportion of instances correctly classified as anomalies relative to the total number of instances that are actually anomalies.
  \item F1-Score: Harmonic mean of precision and recall, providing an overall measure of the model's performance in anomaly detection.
  \item ROC-AUC: The ROC AUC (Receiver Operating Characteristic Area Under the Curve) metric is a performance measure used to evaluate the quality of a binary classification model. The ROC curve is obtained by plotting the true positive rate (TPR or recall) against the false positive rate (FPR or 1 - Specificity) for various threshold values.
\end{itemize}

To evaluate the performance of the anomaly detection algorithms, a comparative analysis was performed between the metrics and the execution time (training or fitting time and testing or data classification time) of the algorithms. The execution time affects the efficiency and scalability of the model. The time required to train and test the algorithms was measured, taking into account the total time for training the models and the average time for classifying each instance.

The analysis of the execution time allows evaluating the practical feasibility of the algorithms in real-world scenarios, especially when dealing with large volumes of data. Algorithms with shorter execution times are preferable, as long as they maintain good performance in the other evaluated metrics. Decreasing the training time, without compromising classification efficiency, is advantageous as it reduces computational resource consumption and training time, especially in applications where there is a constant need for model retraining and evaluation, such as anomaly detection applications, where new intrusion techniques emerge all the time.

On the other hand, decreasing the testing time, without compromising classification efficiency, is equally important in real-time anomaly detection applications, where milliseconds can mean the difference between a successful attack and being blocked by the IDS. The lower the testing time, the higher the probability of blocking malicious activity.

\section{Results and Analysis}\label{sec:resu}

The analysis of the results was divided into (i) analysis of predictive metrics on test sets and (ii) analysis of training times (time required to build the classifier model with the training dataset) and testing times (time required to classify instances in the test dataset).

\subsection{Analysis of Experimentation Scenarios}\label{sec:ana_met}

Tables \ref{tab:resultados_01} and \ref{tab:resultados_02} present the predictive metrics collected from the experimentation scenarios specified in Section \ref{sec:ce}

In these tables, the metrics are rounded to the fourth decimal place for better visualization; therefore, some of the scores displayed as $1.0000$ were actually metrics greater than $0.99995$. However, there were many iterations in which the algorithms correctly classified all instances in the datasets, especially in CE3. It is worth noting that the test data had approximately $50,000$ instances on average, so a model with an accuracy of $0.99995$ misclassified fewer than 3 instances in this data universe, which is a significant result for a classifier. The values of the metrics with the maximum precision obtained by the classifier models can be checked in the project repository.

\begin{table}[ht]
\centering
\caption{Classifier metrics for detection in the first six attacks from the CIC-IDS2018 dataset}
\label{tab:resultados_01}
\includegraphics[width=1\textwidth]{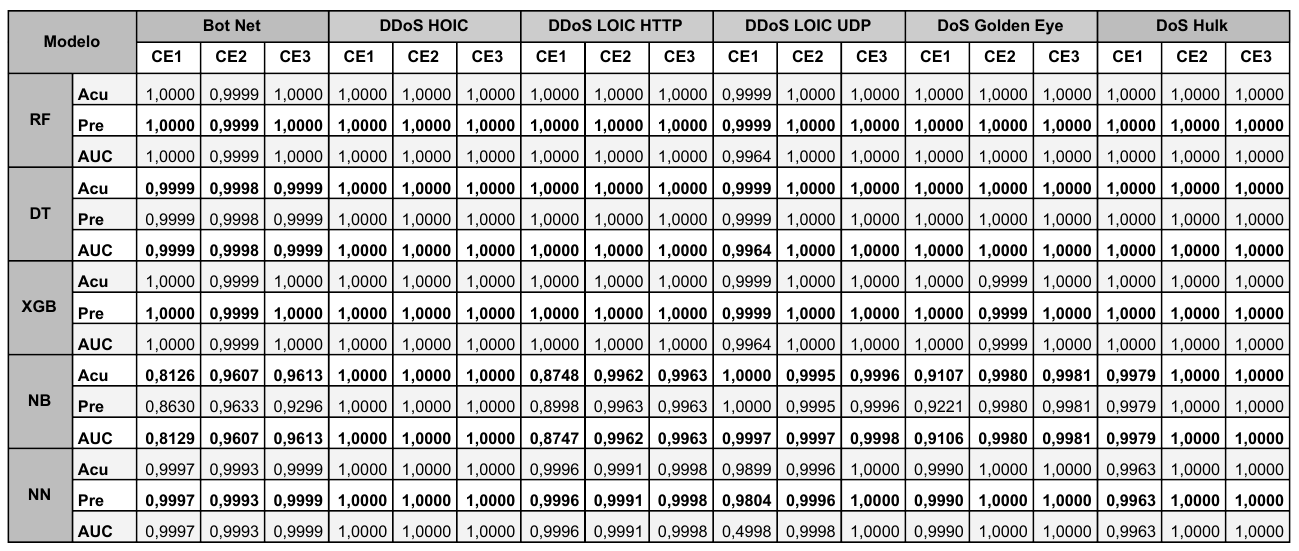}
\end{table}

\begin{table}[ht]
\centering
\caption{Classifier metrics for detection on the last six attacks from the CIC-IDS2018 dataset and the KDD Cup 1999 dataset}
\label{tab:resultados_02}
\includegraphics[width=1\textwidth]{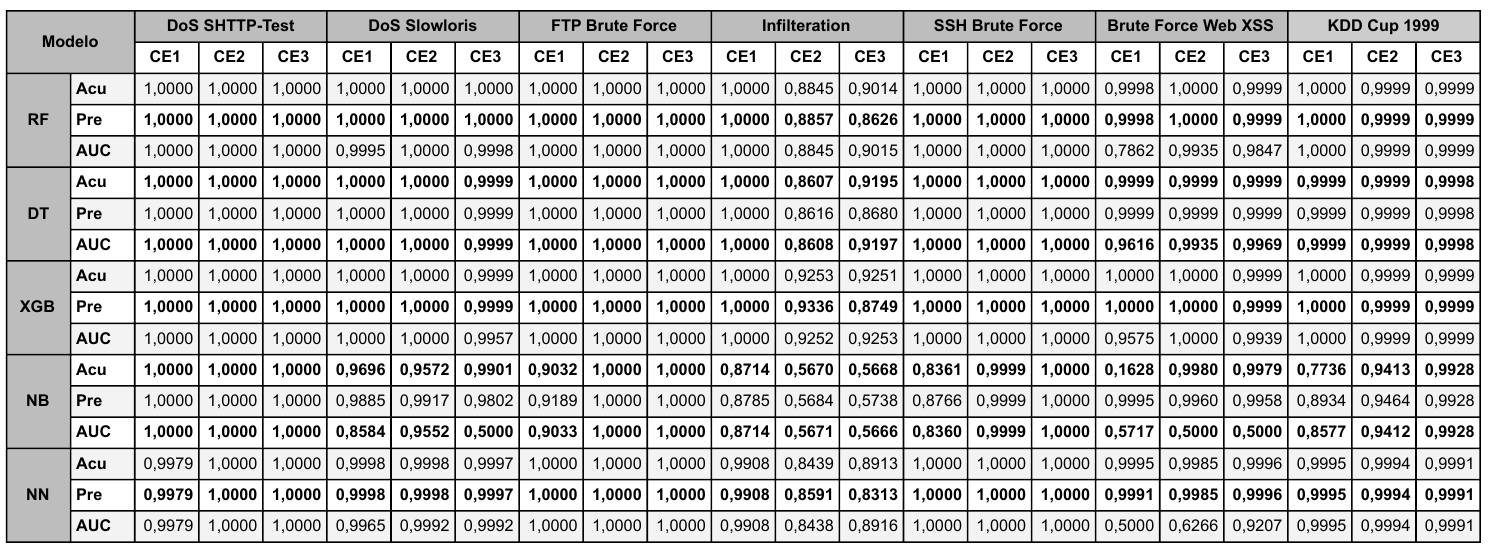}
\end{table}

Comparing the results of the tested ML methods, significant differences in predictive performance can be identified. Random Forest (RF), Decision Tree (DT), and XGBoost (XGB) stood out as the most effective methods, consistently showing high scores in all evaluation metrics, as observed in Tables \ref{tab:resultados_01} and \ref{tab:resultados_02}. In contrast, Naive Bayes (NB) demonstrated less satisfactory performance compared to the other methods, with its scores slightly below the others in terms of accuracy, precision, and ROC-AUC. This can be attributed to its assumption of independence between attributes, which may limit its ability to capture more complex relationships in the data. The Neural Network (NN) presented slightly lower results than the top three methods but still very competitive. The choice of the most suitable method will depend on the specific needs and requirements of the application environment and the dataset used.

Regarding the comparison between the experimentation scenarios, it is noted that, in general, the high scores were maintained even with data and attribute removal in preprocessing, demonstrating that it is possible to maintain and even improve the efficiency of the algorithms in most attacks with preprocessing. Furthermore, NB showed a significant improvement in scores compared to the non-preprocessed dataset. The exception to this behavior occurred in the \textit{Infilteration} attack dataset, where preprocessing decreased the metric scores. In this type of attack, it was observed that it is not beneficial to perform outlier filtering, which may remove relevant instances for training the algorithms in this specific attack with the adopted decision threshold.

The metric used for hyperparameter tuning was accuracy. However, even with improvement in accuracy, there may not necessarily be improvement in all other metrics. It is possible to choose which metrics will be used as a basis for refinement and thus optimize the parameters. This choice depends on the situation in which the modeling is desired to be applied.

\subsection{Analysis of Execution Times}

Table \ref{tab:time_comparison} presents simple arithmetic averages of the execution times (s) of each algorithm on all datasets, followed by a percentage comparison, which allows verification of the decrease in average execution times for each algorithm. The execution times for each type of attack can be checked in the project repository. The execution times for CE3 were not displayed in the table because it would not be a fair comparison with the other times since hyperparameter optimization with cross-validation using the \textit{GridSearchCV} function utilizes parallelization and processing optimization techniques, allowing the use of the machine's processing power more efficiently, resulting in shorter times for each tuning iteration.

\begin{table}[ht]
\centering
\caption{Average execution times of Experimentation Scenarios in seconds}
\label{tab:time_comparison}
\includegraphics[width=1\textwidth]{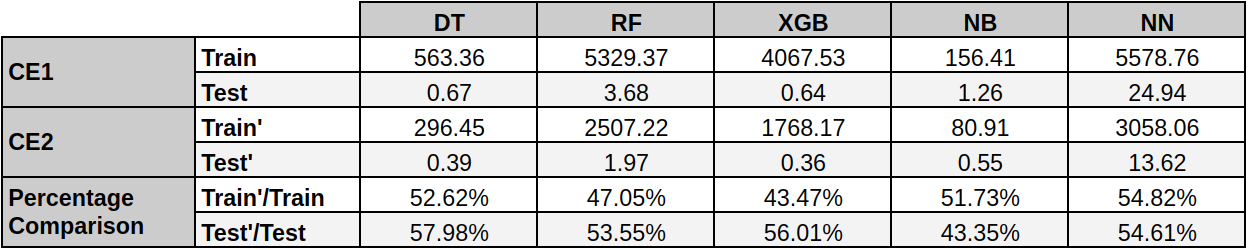}
\end{table}

When analyzing the training and testing times of the analyzed ML methods, significant differences can be identified. Naive Bayes (NB) showed a relatively low tuning time compared to the other methods, indicating efficiency in training the classifier, at the expense of more precise classification, as discussed in Section \ref{sec:ana_met}. This occurs because NB does not train or build the model, it only calculates probabilities from the test dataset. The time for this calculation is referred to in this study, for simplification, as training time.

On the other hand, the RF, XGB, and NN methods showed higher training times, but consistently high predictive metrics. DT presented reduced times and relatively high predictive metrics, thus being a very efficient algorithm for binary classification of the datasets in this study. As for the testing time, it is noted that, except for NN (due to the characteristics of TensorFlow and Keras), all algorithms presented relatively low times. 

Regarding the percentage comparison between CE1 and CE2 times, Table \ref{tab:time_comparison} shows a significant decrease in both testing and training times in all algorithms with data preprocessing, which is relevant for the study in question. It can be noted that in some cases there was a decrease, ranging from 42.02\% to 56.53\%, depending on the algorithm used. The reduction in processing time, with maintenance and improvement in metric scores in most datasets and algorithms, demonstrates that the approach was of great benefit for building the classifier.

It should be noted that the preprocessing time can also be taken into account when comparing testing and training times between CE1 and CE2 since data filtering in CE1 went through fewer processes and thus took less time. The preprocessing times for the CSE-CIC-IDS2018 and KDD1999 datasets are summarized in Table \ref{tab:temp-preproc}.

\begin{table}[ht]
\caption{Preprocessing Times (in seconds)}
\label{tab:temp-preproc}
\centering
\begin{tabular}{|l|c|c|c|c|}
    \hline
    \textbf{Dataset} & \textbf{Outlier Filtering} & \textbf{Normalization} & \textbf{Correlation Filtering} & \textbf{Total} \\
    \hline
    CSE-CIC-IDS2018 & 207 & 34 & 243 & 484 \\
    KDD1999 & 8 & 5 & 17 & 30 \\
    \hline
\end{tabular}
\end{table}

In CSE-CIC-IDS2018, the total preprocessing time for the eleven attacks had a simple average of 44s per attack, much lower than the training time of most algorithms, except for NB, which had an average training time of 57s in CE2. In KDD1999, a greater difference can be observed, with a preprocessing time of 30s against a training time of 371s for NB. Even though there is a similar order of magnitude between preprocessing and training times, NB showed a significant improvement in the analysis metrics. It is worth noting that preprocessing is generally applied only once to the dataset, and with preprocessed data, several training and testing iterations are performed to validate, refine, and retrain the classifier models, making preprocessing time less relevant in a real-world application.

\section{Final Remarks}\label{sec:conc}

Despite the enhancement of IDS with the incorporation of ML techniques for threat identification, the impacts of data preprocessing actions and hyperparameter optimization, as well as their variations, on the performance of classification models and the training and testing execution time of these models have been poorly studied. This work aimed to address the questions associated with the impacts of using these techniques on classifier performance.

The result demonstrated that the use of the preprocessing actions adopted and the optimization of hyperparameter values achieved the best result in the classification performance metrics of most of the evaluated algorithms and attack scenarios, also considerably improving the execution time of both training and testing.

As future work, it is intended to apply such techniques to datasets collected from military networks in training operations to validate the approach's applicability in real-world communication scenarios. Another avenue will be the evaluation of using deep learning algorithms on the collected datasets. Finally, comparing the proposed approach with automation tools for machine learning flow scenarios could be an interesting path to explore.

All resources used in this research, such as complete tables of metrics and execution times for all algorithms, the datasets used, scripts, and programs for automating preprocessing, training, and testing, can be accessed in the project repository (anonymized while the article is in the review phase):

\url{https://github.com/comp-ime-eb-br/ml-4-cyber-def}

\bibliography{bib-template}

\end{document}